# Magnesium-Based Metasurfaces for Dual-Function Switching between Dynamic Holography and Dynamic Color Display

*Jianxiong Li[1,†], Yiqin Chen[2,†], Yueqiang Hu[2,3], Huigao Duan[2,3*], and Na Liu[1,4,5*]*

[1]Max Planck Institute for Intelligent Systems, Heisenbergstrasse 3, 70569 Stuttgart, Germany

[2]State Key Laboratory of Advanced Design and Manufacturing for Vehicle Body, College of Mechanical and Vehicle Engineering, Hunan University, 410082 Changsha, People's Republic of China

[3]Advanced Manufacturing Laboratory of Micro-nano Optical Devices, Shenzhen Research Institute, Hunan University, Shenzhen, 518000, China

[4]Kirchhoff Institute for Physics, University of Heidelberg, Im Neuenheimer Feld 227, 69120 Heidelberg, Germany

[5] Centre for Advanced Materials, University of Heidelberg, Im Neuenheimer Feld 225, 69120 Heidelberg, Germany.

ABSTRACT

Metasurface-based color display and holography have greatly advanced the state of the art display technologies. To further enrich the metasurface functionalities, recently a lot of research endeavors have been made to combine these two display functions within a single device. However, so far such metasurfaces have remained static and lack tunability once the devices are fabricated. In this work, we demonstrate a dynamic dual-function metasurface device at visible frequencies. It allows for switching between dynamic holography and dynamic color display, taking advantage of the reversible phase-transition of magnesium through hydrogenation and dehydrogenation. Spatially-arranged stepwise nanocavity pixels are employed to accurately control the amplitude and phase of light, enabling the generation of high-quality color prints and holograms. Our work represents a paradigm towards compact and multifunctional optical elements for future display technologies.

Metasurfaces, composed of spatially arranged metallic or dielectric antennas, enable a generation of optical elements with diverse capabilities.[1-4] In particular, metasurfaces bring forward interesting design concepts to create light projection and display devices,[5-10] which revolutionize the state of the art display technologies. Different from conventional color generation strategies using pigments and dyes, metasurfaces can yield vibrant and brilliant colors based on resonant excitations of optical antennas[8,9,11,12] or nanocavities.[13,14] The resulting color prints possess excellent stability, long durability, as well as high resolution and high data density, ideally suited for data storage and cryptography applications. Apart from color printing, high-quality holograms can also be produced based on metasurfaces.[6,7,15-17] Such optical devices are ultra-thin and ultra-compact, constituting a milestone for modern display technologies.

Often, color printing and holography are separately implemented on metasurface devices. Very recently, research endeavors have been exerted to combine these two display functions within a single metasurface device.[18-22] For example, monolithic stepwise nanocavity and holographic metasurface layers have been stacked together to achieve low-crosstalk color printing and full-color holography.[19] In addition, single-layer dielectric metasurfaces have been employed to combine color printing and holography.[18,20,22] However, these two strategies suffer from low efficiency and polarization sensitivity, respectively. More crucially, these metasurfaces are intrinsically static, as their optical functions are fixed once the devices are fabricated. This leaves out many opportunities that metasurfaces can offer.

In this work, we demonstrate a metasurface device, which allows for dual-function switching between dynamic holography and dynamic color display. The dynamic functionality is enabled by the phase-transition of magnesium (Mg), which can be transformed to magnesium hydride ($MgH_2$) through hydrogen uptake.[23-30] Importantly, $MgH_2$ can be restored to Mg upon oxygen exposure. Such reversible metal to dielectric transitions are accompanied with dramatic optical response changes, which lay the foundation for building dynamic optical devices in the visible wavelength regime.

**RESULTS AND DISCUSSION**

Figure 1a shows the schematic of the metasurface, which consists of spatially arranged stepwise pixels of subwavelength dimension. The metasurface can work as a holography device, creating high-quality holographic images. The metasurface can also work as a color display device, which produces microprints with vivid colors. These two display functions, dynamic holography and dynamic color display, can be readily switched through hydrogenation and dehydrogenation using hydrogen and oxygen, respectively. Fig. 1b illustrates the fabrication process of the dual-function metasurface, which consists of stepwise pixels. Each pixel is a Fabry−Pérot (FP) nanocavity of subwavelength lateral dimension (500 nm). A silver (Ag, 100 nm) film and a silicon dioxide ($SiO_2$, 190 nm) film are deposited on a silicon substrate, respectively. Here, the $SiO_2$ layer is utilized to simplify the fabrication process, as it can contribute a common height for all the nanocavities. Stepwise pillars, made of hydrogen silsesquioxane (HSQ) are fabricated through multiple overlay procedures by electron beam lithography with the help of gold alignment markers. Subsequently, Mg (50 nm)/titanium (Ti, 2 nm)/palladium (Pd, 3 nm) films are successively deposited on the patterned sample.

To elucidate the working principle of the dual-function dynamic metasurface, Fig. 2a shows the schematic diagrams of light propagation in a nanocavity pixel at the Mg and $MgH_2$ states, respectively. At the Mg state, the metallic (Mg/Ti/Pd) capping layers efficiently reflect the incident light in the visible wavelength range. The phase of the reflected light can be modulated by varying the HSQ height ($t_{HSQ}$). Four different HSQ heights, $t_{HSQ}$ = 0, 80, 160, and 240 nm (see black stars in Fig. 2b) are employed to define the discrete phase levels spanning in a $2\pi$ range at 633 nm, while maintaining high reflectance. The light wavefronts can thus be well controlled to generate phase-only holograms in the far field. Upon hydrogen exposure, Pd catalyzes the dissociation of hydrogen molecules into hydrogen atoms, which diffuse through Ti into Mg. Mg undergoes the metal to dielectric transition to form $MgH_2$. The incident light can then transmit through the capping layers and experience multiple reflections inside the FP nanocavity,[31] which is formed between the capping layers ($TiH_2$/$PdH$) and the bottom Ag mirror (see Fig. 2a). In this case, the cavity height is defined by the total thickness of $MgH_2$ + HSQ + $SiO_2$. As shown in the upper panel in Fig. 2b, in contrast to the nearly unity reflectance for different pixels before hydrogenation (black curve), the reflectance intensity exhibits a clear dependence on the cavity height after hydrogenation (red curve). In addition, the associated phase differences become much smaller among different pixels (see red stars in Fig. 2b), when compared to those before hydrogenation.

A holographic image of a musical note is designed based on the Gerchberg-Saxton algorithm and its phase distribution is shown in Fig. 2c. Fig. 2d presents the scanning electron microscopy (SEM) image of the fabricated sample. A tilted view is shown as inset in Fig. 2d. A high-quality hologram of a musical note with 58% efficiency is experimentally reconstructed at 633 nm (see Fig. 2e), demonstrating the precise control of the stepwise pixel heights and thus the phase distribution. The incident laser light is linearly polarized along one lateral side of the stepwise pixels in the experiment. However, the reconstruction of the hologram is independent on the light polarization. It is noteworthy that the hologram is detected at an 11.5° off-axis angle along only one direction. Upon hydrogen exposure (5% in nitrogen), Mg is transformed to $MgH_2$. The phase differences among different stepwise pixels significantly decrease, resulting in the disappearance of the hologram. When $MgH_2$ is restored to Mg upon oxygen exposure (10 % in nitrogen), the hologram comes into existence again. As a result, the phase-transition between Mg and $MgH_2$ enables dynamic holography that can be reversibly switched on/off by loading $O_2$/$H_2$.

Meanwhile, through hydrogenation and dehydrogenation, the stepwise pixels also experience abrupt changes in reflectance, rendering dynamic color generation possible. As revealed by the optical microscopy images in Fig. 3a, before hydrogenation the pixels with different HSQ heights exhibit blank color when interacting with unpolarized white light, owing to their high reflectance in the visible wavelength range. This is also confirmed by the measured reflectance spectra of these pixels, which are characterized by the black curves in Fig. 3b. After hydrogenation, the formation of the sharp FP resonances greatly modulate the reflectance spectra (red curves), resulting in vibrant color generations from the stepwise pixels (see Fig. 3a). The FP resonance positions are largely dependent on the cavity height. The resonances shift to longer wavelengths, when the cavity height increases. Figs. 3c and 3d present the simulated reflectance spectra and the electric field distributions, respectively, for a representative case of $t_{HSQ}$ = 240 nm. The two reflectance peaks at 755 nm and 510 nm correspond to the formation of the second- and third-order FP resonances, respectively (see the red curves). After dehydrogenation, the stepwise pixels return to the blank color state again. Therefore, the phase-transition between Mg and $MgH_2$ also enables dynamic color display that can be reversibly switched on/off by loading $H_2/O_2$.

The two different display functions, dynamic holography and dynamic color display, can be integrated within one metasurface device for dual-function switching. The SEM image of the dual-function metasurface is presented in Fig. 4a. An enlarged view is shown as inset in the same figure. Stepwise pixels with four different HSQ heights are employed on the metasuface to create a color print ' $E = h\nu$ ' and a holographic image of the Max-Planck Society logo. Fig. 4b shows the experimental results. Before hydrogenation, under illumination of white light the metasurface exhibits blank color, whereas at the incidence of laser light at 633 nm the metasurface generates a hologram of the Max-Planck Society logo. Upon hydrogen exposure, colors start to appear under white light illumination and a color print of '$E = h\nu$' becomes vividly observable. The greenish background is due to the uniform superposition of the colors generated from the four sets of stepwise pixels that are used to construct the hologram. On the other side, after hydrogenation the hologram of the Max-Planck Society logo disappears. Subsequently, upon oxygen exposure, the color print vanishes and the hologram occurs again. Fig. 4c shows the time evolutions of the gray level of the color print and the hologram intensity during hydrogenation and dehydrogenation, respectively. The evolution of the gray level, which indicates the pixel brightness changes, is obtained by tracking the averaged gray values of the area with '=' pattern in Fig. 4a. It decreases upon hydrogen exposure, when the print transits from the blank state to the color state. It then increases upon oxygen exposure, when returning from the color state to the blank state. Meanwhile, the light intensity of the hologram decreases and subsequently increases, when the hologram is switched off/on through hydrogenation/dehydrogenation, respectively. All the experiments have been carried out at 80 °C to facilitate the switching upon $H_2$ and $O_2$ loading. The operations of the dynamic color display and dynamic holography are shown in Supporting Movies S1 and S2, respectively.

**CONCLUSIONS**

In conclusion, we have realized a dual-function metasurface for switching between dynamic holography and dynamic color display. The spatially-arranged stepwise nanocavity pixels allow for accurate control over the amplitude and phase of light, enabling the facile design of metasurface color prints and holograms. In addition, the reversible phase-transition between Mg and $MgH_2$ enables the dynamic functionality, leading to switchable holography and color display at visible frequencies. Our work represents a paradigm to realize multi-tasking optical elements, which broads the realm of light projection devices and display technologies.

**EXPERIMENTAL METHODS**

**Sample fabrications**

The substrate was prepared by electron beam evaporation of Ag (120 nm) and SiO$_2$ (190 nm) on a Si wafer. The multiple overlay patterning of the stepwise metasurface was performed using a Raith electron-beam lithography system with an accelerating voltage of 30 kV and beam current of 890 pA. The spin-coated HSQ film was directly utilized without prebaking to avoid thermally induced cross-linking. Before carrying out the overlay fabrication, gold alignment markers were defined by a standard "Sketch and Peel" lithography process. To exactly define the heights of the stepwise pillars, the spinning speed and the concentration of the HSQ resist were finely tuned in each overlay. The details of the spin coating and exposures are shown in Table S1. The development time was 15 s to avoid the damage of the pillars fabricated in the previous overlays. Subsequently, Mg (50 nm)/Ti (2 nm)/Pd (3 nm) films were successively deposited on the patterned sample by electron beam evaporation.

**Optical setups**

The holograms were recorded by an optical setup shown in Figure S1. The light beam was generated from a laser diode source (633 nm). A linear polarizer (LP) was employed to obtain linearly polarized light. An optical lens and an aperture were utilized to reshape the light beam to a similar size as the sample area in order to avoid undesired reflected light. The light beam was incident on the metasurface sample placed in a homemade gas cell. The reflective hologram was projected onto the screen in the far field. All the experiments were carried out at 80°C to facilitate the switching upon H$_2$ or O$_2$ loading. The hydrogenation and dehydrogenation experiments were carried out in a homebuilt stainless steel chamber. Ultrahigh-purity hydrogen, oxygen, and nitrogen were used with mass flow controllers to adjust the flow rates and gas concentrations in the chamber. The flow rate of the hydrogen and oxygen gases was both 1.0 L/min.

The color microprint was revealed using a NT&C bright-field reflection microscopy set-up (using a Nikon ECLIPSE LV100ND microscope) illuminated by a white light source (Energetiq

Laser-Driven Light Source, EQ-99). A digital CCD Camera (Allied-Vision Prosilica GT2450C) was used to capture the color micrographs with a 10× (NA = 0.3) objective. The optical spectra were measured in a reflection mode using a microspectrometer (Princeton Instruments, Acton SP-2356 Spectrograph with Pixis:256E silicon CCD camera) with the electric field of the unpolarized light in plane with the substrate surface. The reflectance spectra were measured from samples in a stainless-steel cell during hydrogenation and dehydrogenation. The measured reflectance spectra were normalized with respect to that of an Al mirror.

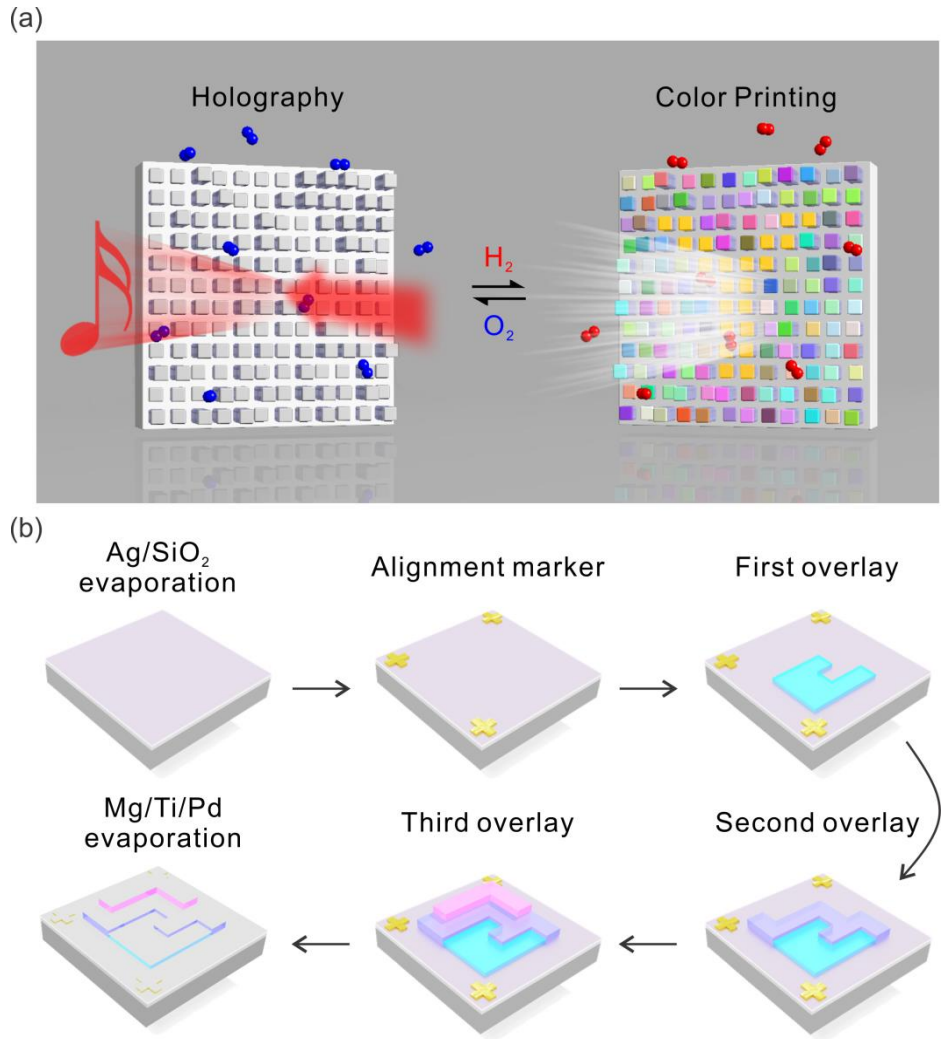

**Figure 1.** (a) Schematic illustration of the dual-function switching between dynamic holography and dynamic color display by hydrogenation ($H_2$) and dehydrogenation ($O_2$). (b) Fabrication process of the dual-function metasurface device.

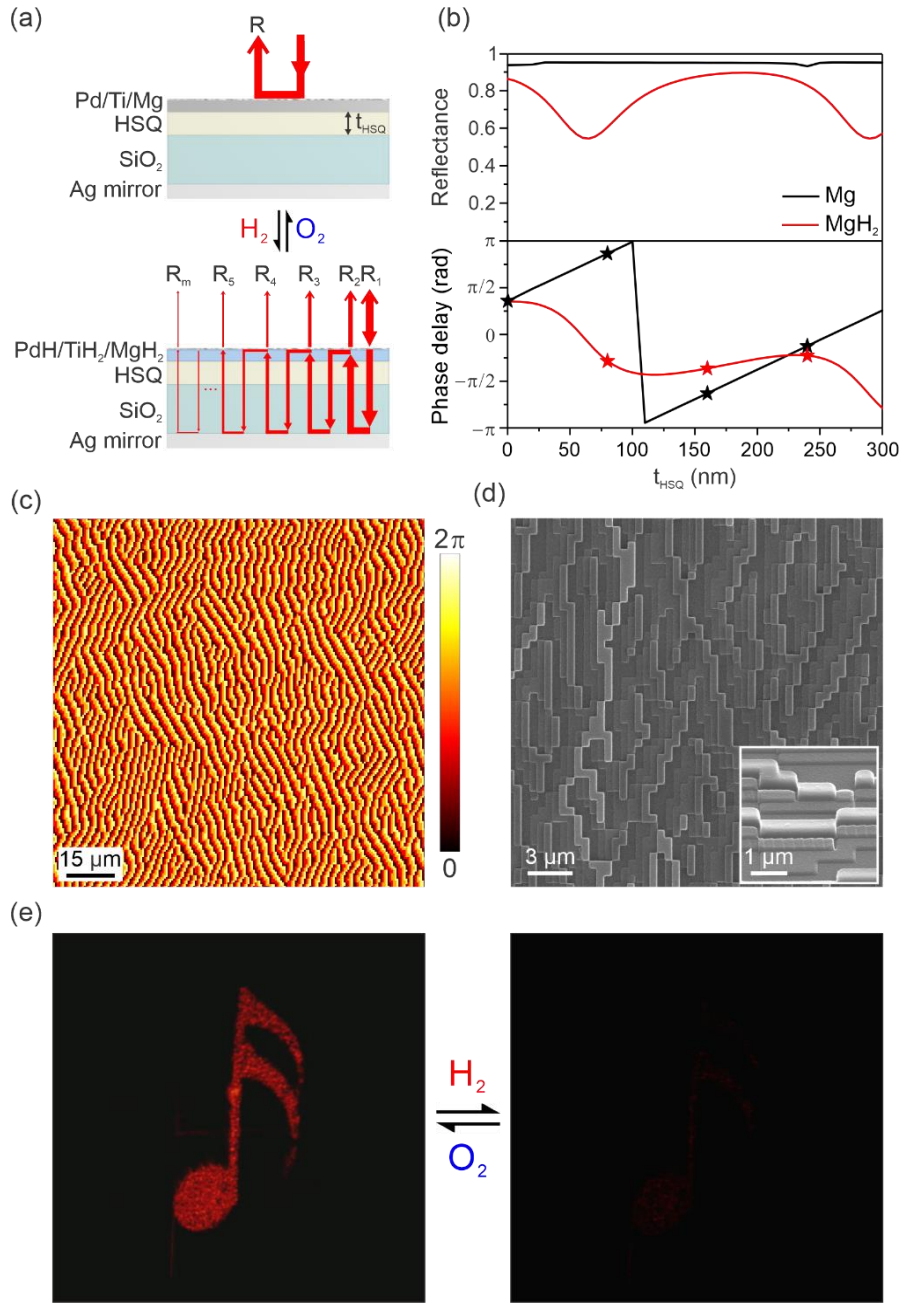

**Figure 2.** Working principle and experimental results of dynamic holography. (a) Schematic diagrams of light propagation in the metasurface device before and after hydrogenation. (b) Simulated reflectance spectra and phase delay with incident light at the wavelength of 633 nm at the Mg (black) and $MgH_2$ (red) states. (c) Phase distribution for the hologram of a musical note. (d) SEM images of the metasurface sample. (e) Experimentally recorded switchable hologram through hydrogenation ($H_2$) and dehydrogenation ($O_2$).

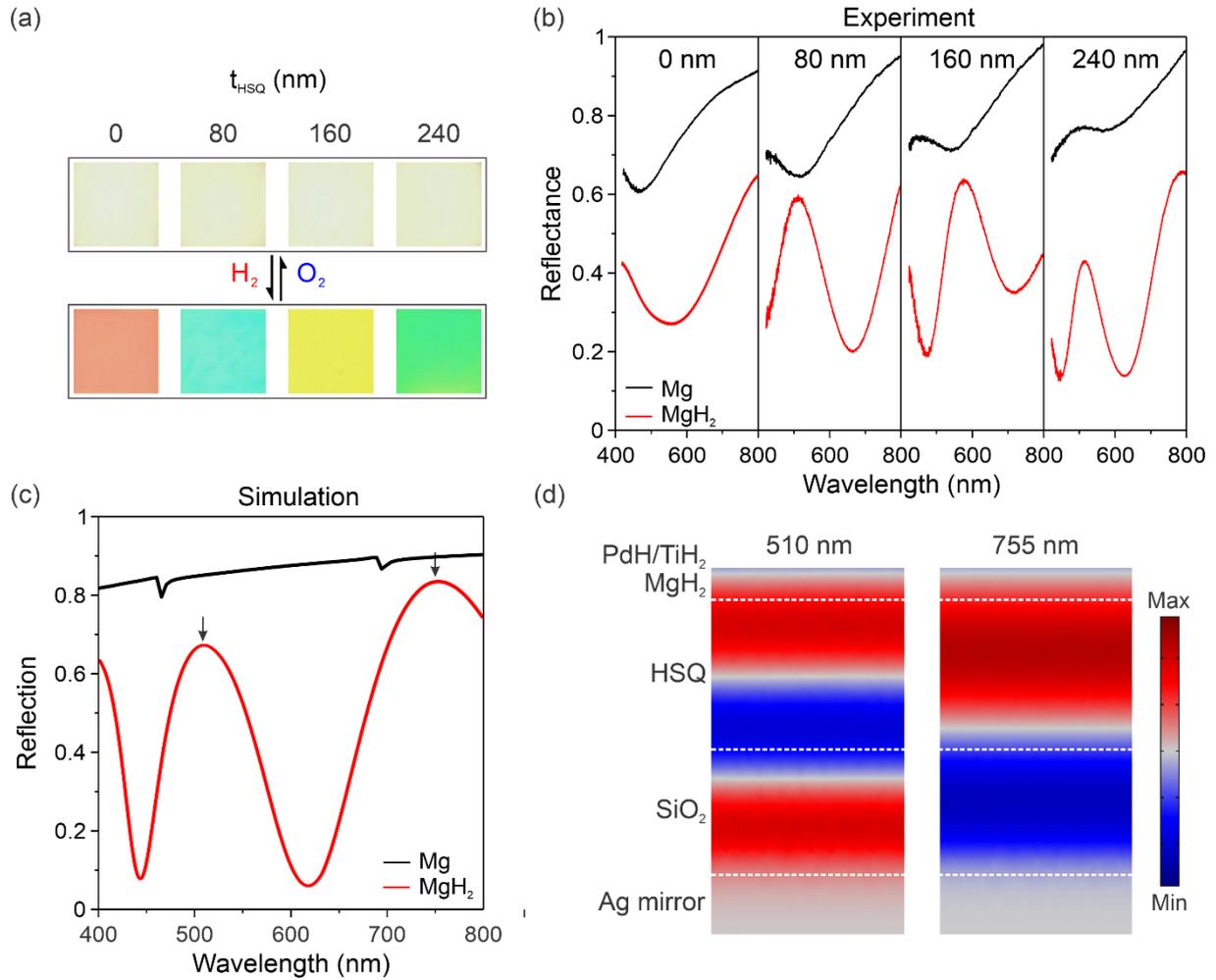

**Figure 3.** Experimental results of dynamic color display. (a) Optical microscopy images and (b) experimental reflectance spectra of the stepwise pixels with different heights, $t_{HSQ}$ = 0, 80, 160 and 240 nm before and after hydrogenation. (c) Simulated reflectance spectra of the pixel ($t_{HSQ}$ = 240 nm) before and after hydrogenation. (d) Electric field distributions at the different FP resonances (see the black arrows in (c)) after hydrogenation. Third- and second-order FP resonances are excited at 510 nm and 755 nm, respectively.

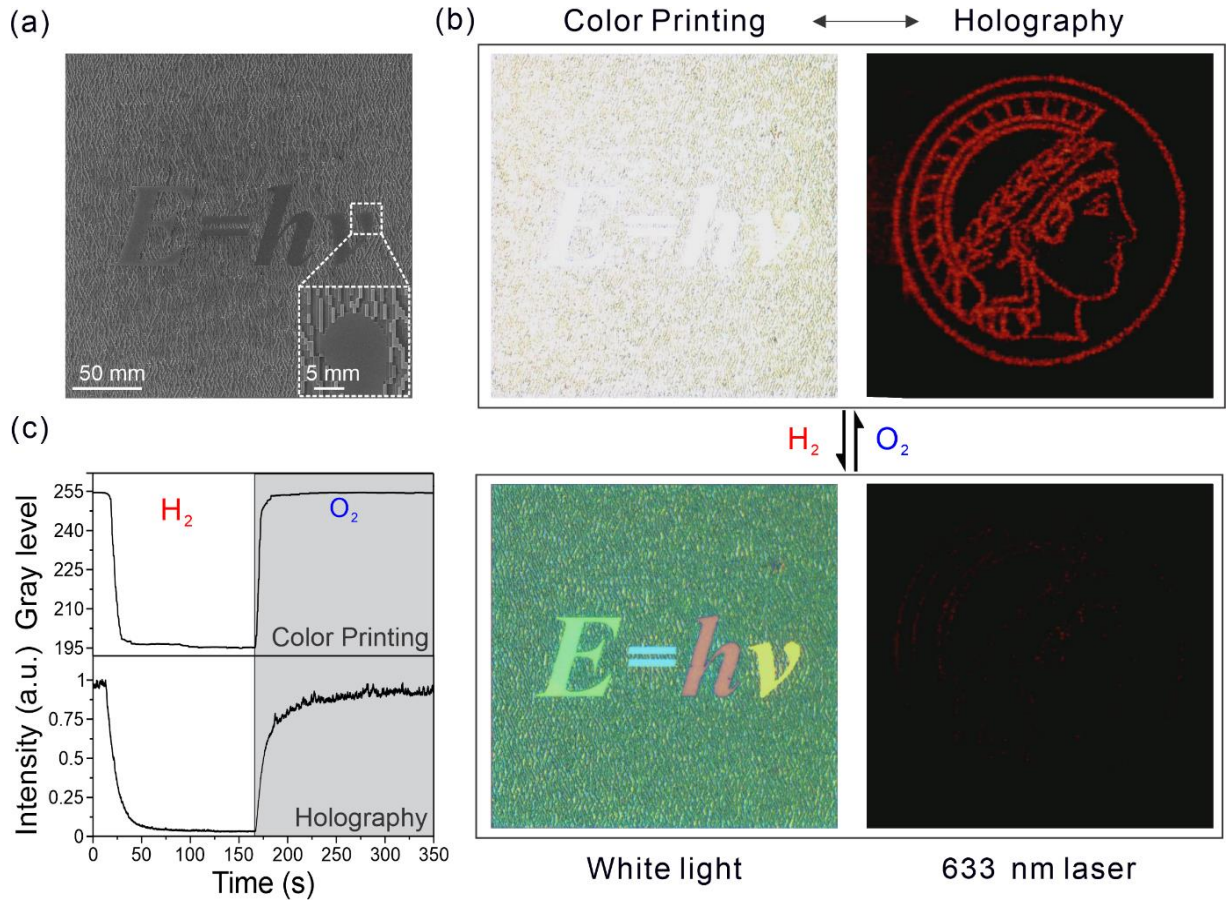

**Figure 4.** Dynamic metasurface with dual-function. (a) SEM images of the dual-function metasurface. (b) Switching between dynamic color display and dynamic holography through hydrogenation and dehydrogenation. The Max Planck society logo is used with permission. (c) Time evolutions of the gray value of the color print and the hologram intensity during hydrogenation and dehydrogenation.

## ASSOCIATED CONTENT

**Supporting Information.**

The following files are available free of charge.

Supporting information about numerical simulation parameters and holography designs. (word)

Movie S1. Video that records dynamic color display. (MP4)

Movie S2. Video that records dynamic holography. (MP4)

## AUTHOR INFORMATION

**Corresponding Author**

*Email: duanhg@hnu.edu.cn

*Email: na.liu@kip.uni-heidelberg.de

**Author Contributions**

† Both authors contributed equally to this work. The manuscript was written through contributions of all authors. All authors have given approval to the final version of the manuscript.

## ACKNOWLEDGMENT

This project was supported by the European Research Council (ERC Dynamic Nano) grant as well as the National Natural Science Foundation of China (Grants 51722503).The authors thank the support from the Max-Planck Institute for Solid State Research for the usage of clean room facilities.

ToC Figure:

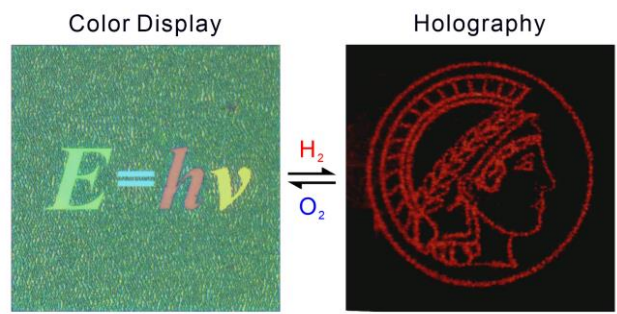